# Unraveling Specific Conditions for a Repeatable Mpemba Effect

*James D. Brownridge[a], Matthieu Zinet[b], Paul Sotta[c] and Francois Ganachaud*[c,d,e]*

[a] Physics Department, Binghamton University, 4400 Vestal Parkway East, Binghamton, NY 13902, USA.

[b] Engineering of Polymer materials, IMP, UMR5223 CNRS-INSA Lyon-UCBL-UJM, Bâtiment Polytech, 15 Bd Latarjet, 69622 Villeurbanne Cedex, France

[c] Engineering of Polymer materials, IMP, UMR5223 CNRS-INSA Lyon-UCBL-UJM, Batiment Jules Verne, 17 Avenue Jean Capelle, 69623 Villeurbanne Cedex, France.

[d] Complex Assembly of Soft Matter, COMPASS, UMI3254 CNRS-Solvay-Upenn, 350 Georges Patterson Boulevard, Bristol PA 19007, USA

[e] Bicarlab, 31 rue Octave Mirbeau, 69150 Décines Charpieu, France

ABSTRACT. Water exhibits many unique properties compared to other liquids, with some of these explained and others remaining enigmatic. Among them, it was proposed and extensively debated that hot water would freeze faster than cold water. Numerous studies have demonstrated the difficulty of successfully elucidating this effect, making explanations surrounding this phenomenon highly controversial. Here, we demonstrate that when two cups filled with cold and hot water are introduced simultaneously in a freezer saturated with ice-nucleating agents, the hot sample freezes faster and to a greater depth than the cold sample, particularly when the initial temperature difference is high. On the other hand, against some previous beliefs, the time to onset of crystallization is always and logically retarded for hotter samples. Under these conditions, where supercooling is eliminated and temperature recording is precisely controlled, robust experiments follow the same trend, regardless of whether hot versus room temperature (RT) samples or RT versus cold samples are tested. Differences in heat transfer are proposed and simulated to explain such divergence in freezing time in compliance with Newton's law. This work confirms the original study of Mpemba and Osborne, whose results have been so difficult to replicate, without questioning the burgeoning research on related effects.

## INTRODUCTION

Water is a unique substance that provides essential services to the development of life and humanity and possess extraordinarily complex physical properties. For instance, in his extensively documented website 'water structure and science'[1], Martin Chaplin has listed several physical abnormalities of water (10 items to date, e.g., density, viscosity), phase anomalies (13 items, e.g., high boiling and melting points) or unexpected properties (6 items, e.g., the temperature dependence of most physical properties). Water research is a vibrant domain that generates considerable excitement (see recent editions of Chemical Reviews on the subject[2,3]). A vast community of researchers is very active in this field, trying to explain strange effects that have been around for years (e.g., the Ray-Jones effect[4,5] or the Hoffmeister series[6]) or coming up with new notions to account for unexplained phenomena, for instance, around surface tension (e.g., nanobubbles at colloidal interfaces[7]).

The Mpemba effect, which stipulates that 'hot water freezes faster than cold' (see a now old review here[8]), has been another long-lasting debate in the Soft Matter domain. This effect was studied by Mpemba and Osborne and published in 1969[9]. In subsequent years, various teams have attempted to reproduce these experiments with limited success (a series of papers came out in the same journal in a report-comment like manner[10,11,12,13]; see also[14,15]). In the 90s-00s, thorough studies using adapted devices and pure water were published from time to time, all of them proposing different explanations for this effect (preferential evaporation of hot water, influence of dissolved gases in cold water, enhanced convection in hot water, or influence of the surrounding environment, e.g., beaker support….) and relatively poor experimental reproducibility, if any at all[16,17,18,19,20,21,22]. Things have become even more confusing recently, with several articles reanalyzing different data and reviving the controversy (see, e.g., two recent analytically reviewing

papers[23,24]). Additionally, several theoretical articles have proposed that the Mpemba effect can be attributed to the specific physics of water molecules: the H-bonding network would organize differently as a function of temperature, thereby favoring faster crystallization[25,26]. The only proven Mpemba effect published to date was conducted at the colloidal scale, with the "hot" initial condition equilibrating up to ten times quicker than the "cold" one (albeit with a freezing time difference of only a few milliseconds)[27].

No clear explanation nowadays wins unanimous support, ultimately questioning the veracity or even the actual existence of the Mpemba effect[28]. This may be ascribed in part to the lack of definition of the process, as mentioned earlier[23]: i) is it the onset of freezing or the total freezing time that should be considered?; ii) which temperature difference should be implemented? Besides, experimental discrepancies prevent reproducibility. One is the location of the temperature probe that conditions the measured total freezing time. Another key issue is supercooling: it is nearly impossible to reproduce the same conditions of ice nucleation from one vessel to another, and even sometimes in the same vessel (see, e.g.[29]). Some authors have recently suggested that the Mpemba effect might arise from the stochastic nature of heterogeneous ice crystal nucleation at the water/container interface[30]. They performed careful experiments to demonstrate this stochastic behavior. They observed significant variations in freezing times, defined as the onset of freezing, which led to a sharp temperature increase due to latent heat release. This is mainly because of the highly non-linear relationship between nucleation time and the supercooling increment, $\Delta T$. However, in the macroscopic systems considered here, thermodynamic fluctuations are negligible[31]. The entire experimental setup and the macroscopic thermal processes, rather than individual microscopic relaxation events, must be taken into account[32].

An extension of the notion of Mpemba effect has emerged more recently[33]. In a review recently posted on ArXiv[34], some authors expand the idea of the Mpemba effect to include a series of observations of Mpemba-like effects, in which various physical systems reach equilibrium faster when the initial distance to equilibrium is larger, in apparent contradiction with Newton's Law of Cooling. Besides freezing water, examples include magnetic alloys, spin glass systems, and other systems. The authors attempt to categorize reported 'Mpemba-like effects' based on the nature of the initial and final states (whether these are actual thermal equilibrium distributions or non-equilibrium steady states) and on the type of variation in the relaxation time as a function of the initial and final states. Simulations have since been extended to gases and granular media, and numerous articles currently discuss the 'quantum Mpemba effect.' It is essential to note that relaxation far from equilibrium should, of course, be first described according to general frameworks of non-equilibrium thermodynamics,[35,36] even if it does not offer an insight into particular physical mechanisms that drive the evolution of a specific system in a given experimental situation.

In the following, we stick to the original phenomenology of the Mpemba effect, namely the freezing of macroscopic cups of water and other liquids. Upon examining the original conditions of Mpemba's work, we identified two essential features that had not been systematically considered. First, the significant content of ice crystals surrounding the samples in a commercial freezer allows very efficient heterogeneous nucleation, thus preventing supercooling. Second, all experiments reported in the original paper compared hot and cold water frozen simultaneously in the same device. We implemented these conditions to allow reproducible measurements and precise analyses. Accordingly, we define the effect as the difference in freezing times between hot and cold cups. This time is precisely recorded for each cup from ice nucleation (at T = 0 °C) to complete freezing (when T passes below -1 °C). We first describe the experimental setup and the procedure to record the time to onset of freezing and the total freezing time of each cup while avoiding supercooling. We then analyze these data in light of previous hypotheses and complementary experiments to strengthen the plot. In the second part, thermal exchanges in the device are simulated, enlightening the importance of heat transfer at each sample's freezing time. A final discussion wraps up the study.

## EXPERIMENTAL SECTION

### Experimental device

An experimental station (Fig. 1) was designed and constructed to produce consistent and reproducible results over time. The station consisted of a top-loading chest freezer with a volume of 150 $dm^3$ and a minimum temperature of approximately -30.5 °C. A 10 cm diameter hole was cut at the top, above where the test water samples were to be placed. The hole was covered with a metal screen that permitted the introduction of ice crystals. These would be produced when moisture in the room air came into contact with a liquid nitrogen-filled styrofoam cup seating on the screen and penetrated through perforated ice nucleators onto the surface of the water in each of the two cups. The water samples were held in plastic cups. The cups were held by their rims around the tops and suspended ~20 cm below the hole in the top of the freezer. It was essential that the parts of the cups in contact with the water not come into contact with any solid object. We used a two-cup system with the distance between the cups fixed at 7cm from center to center. We used Type K thermocouples to measure the temperatures and a Measurement Computer DAQ USB-2408 unit to collect and record the temperature every 3.2 seconds as the water cooled, froze, and thawed. Each cooling, freezing, and thawing cycle lasts ~24 hours. The position of the thermocouples tips was also fixed so they would not move between freezer thawing cycles. We had previously determined that it is essential for the tips of the thermocouples to be positioned and held in place just under the water's surface, in the center of each cup, to determine the freezing time accurately. The freezer and DAQ were located in a temperature-controlled room, with the expectation that it would take several months to conduct the necessary set of experiments based on previous work.

### Computer simulations

The finite element mesh for our simulations was generated using a free triangular meshing algorithm. The order of the shape function describing the temperature field within each element is linear, providing a balance between computational efficiency and accuracy in capturing temperature variations. The mesh size was adjusted independently for each domain to represent the temperature field accurately. Specifically, a maximum element size of 1 mm was allowed for the water domains. The average number of elements in the two water domains is 3210. The discretization of the air domain involves elements with progressively larger sizes as the distance from the cups increases due to an expected lower temperature gradient in those areas. The resulting mesh comprises a total of 17,165 elements. We have verified that the mesh refinement is sufficient to avoid introducing significant biases in the presented results.

The time-dependent solution used the linear direct solver PARDISO (Parallel Direct Sparse Solver) and the Newton-Raphson method for nonlinear iterations, incorporating a constant damping factor of 0.9. A relative tolerance of 10-5 was used for all simulations. To optimize the simulation time, the requested output time step ranged from 0.1 to 1 minute based on the progression of the simulated cooling process. Precisely, freezing onset and end times were determined with a resolution of 0.25 min. However, the solver dynamically adjusts the calculation time step based on the convergence of the solution to the set of equations, considering the fixed tolerance. This resulted in the actual time step taken by the solver varying from $5x10^{-6}$ s at the start of the simulation to 60 s during the final cooling phase after freezing. The calculations were performed on a 13th-generation Intel Core i7 Windows PC with 16 GB of RAM. The average CPU time required to solve each simulated case was approximately 2 minutes.

## RESULTS

### Experimental setup and robustness of the measurements

Reproducing the Mpemba effect with the utmost meticulousness at the lab scale proved to be challenging. Our device is a conventional freezer where two beakers filled with the same liquid (typically distilled water) at different initial temperatures are suspended and gradually cooled down to -30°C, the freezer's temperature. One beaker is the reference cup (later referred to as RC) set at 20°C, while the second test cup (later referred to as TC) is set at a different temperature. Figures 1a and 1b show the schematics and photos of the device (see also the experimental part for more technical details). Different features are essential to insure perfect reproducibility of the experiment. First, the cups are suspended to prevent heat exchange with the support. Indeed, one previous explanation of the Mpemba effect[8] relied on the fact that hot water could melt the ice on the support where the cup sits and then accelerate freezing through better thermal conduction within the cup. This cannot happen here. Second, a Styrofoam cup filled with liquid nitrogen was placed above a hole on top of the device. The ambient humidity generates large quantities of ice flakes in the surrounding atmosphere of the beakers, ensuring efficient ice nucleation and preventing supercooling. Finally, it is known from one of the authors' previous study[20] that the location of the temperature probe heavily conditions the results (Figure 1c). We have placed the thermocouples at the centers of the beakers top to record the precise time of freezing. The fact that the probe is precisely located and remains tightly in place throughout the experiment is crucial for reproducibility. As water has its maximum density at 4°C, the entire cup reaches 0°C simultaneously. Indeed, whereas the liquid at the bottom of the cup rapidly freezes and thus decreases in temperature slowly over time, the top probe remains at 0°C until the entire sample is frozen.

To demonstrate that reproducibility in measurements can be achieved with such a complex liquid as distilled water, blank experiments with the same initial temperature in both cups were systematically conducted before proceeding. Four different experiments, initiated at room temperature (RT), are shown in Figure 1d. Comparing blank experiments beginning at RT and at 56°C (Figure S1), the freezing time is the same (117 and 115 min, respectively). The absence of differences in freezing times when cups are initially set at the same temperature has been verified multiple times (not shown). In few instances, we have observed that at high temperatures (typically above 60°C), mild supercooling of about -0.5 to -1°C occurred (see an example in Figure S1, blank trial started at 64°C). In that case, the latent heat jump decreases the total freezing time. This relatively small supercooling was typically observed when the initial temperature of the cup was hot; this may partly explain the discrepancy in the data observed for some trials, even though we did not explore this possibility. For the record also, we have checked, using a very hot TC sample (73°C), that the loss of water during a trial is certainly not negligible but relatively equal between each cup (starting from 53 g, we went down to 46.7 and 49.4 g for hot and RT cups, respectively). This result rules out evaporation as an explanation for the Mpemba effect. Other conditions (nature of water, cup swapping, and water content variations) were also examined, as summarized in the supporting information.

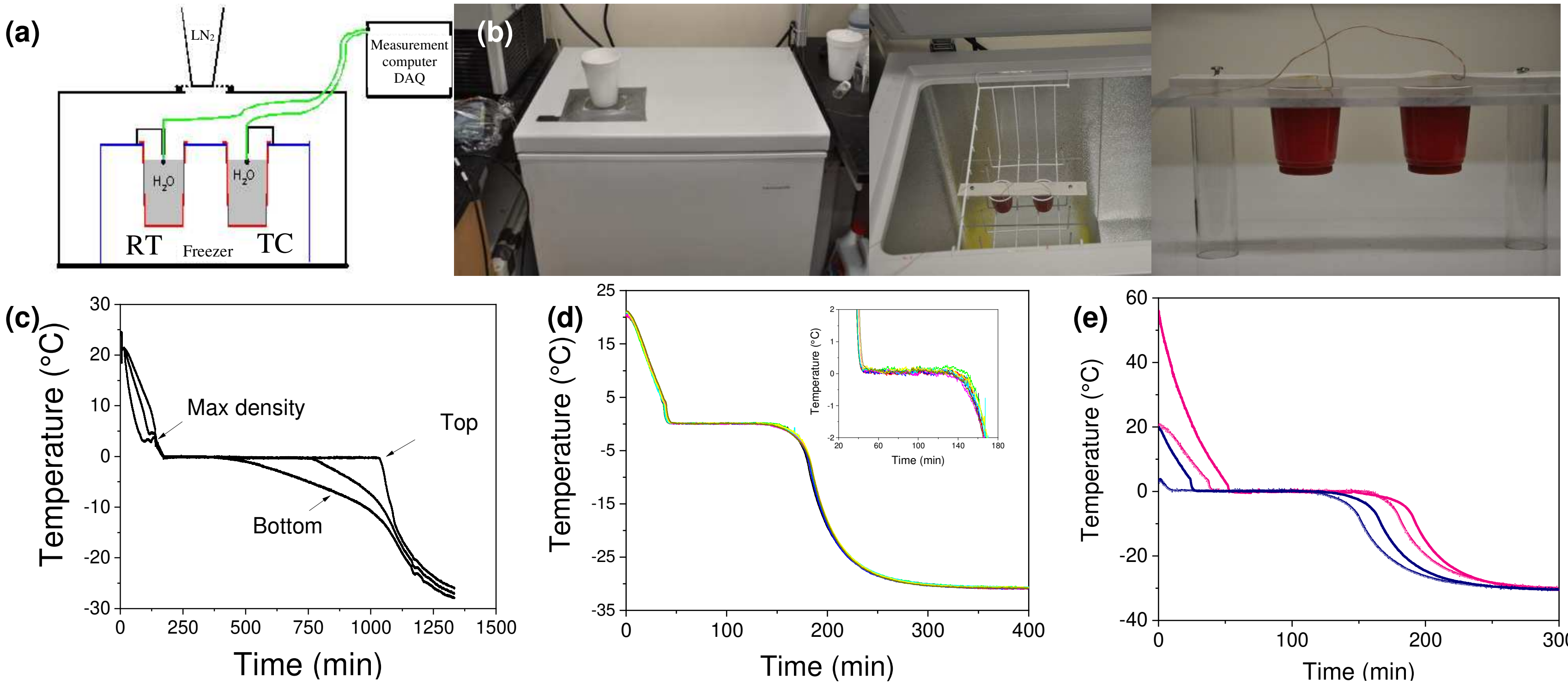

**Figure 1 – Set-up and robustness of measurements**. **(a)** Schematics of the device used in this study. **(b)** Experimental device: the freezer is mounted with a hole on which a liquid nitrogen cup releases ice from ambient humidity. Inside the box, two plastic cups containing approximately 50 g of water are frozen and monitored by thermocouples placed at the exact center and top of the liquid. Thermocouples record the temperature in each cup as a function of time. **(c)** Influence of the position of the temperature probe on the estimation of the freezing time; **(d)** Typical thermograms for eight samples, all frozen from RT to show the reproducibility of the experiments (inset: zoom on the freezing step; see also Figure S1). **(e)** Typical measurements carried out in this study, starting from pre-thermalized TC at temperatures lower (4°C versus 21 °C, blue curves) or higher (55°C versus RT, in pink) than RT.

Typical experiments carried out in this study are given in Figure 1e. We both conducted experiments with TC containing hot or cold (pre-cooled) water. In the figure, two experiments are shown: one where the TC starts at 4 °C and the other one where the TC cup is heated up to 54°C before introduction in the freezer. These experiments were reproduced many times. Only one cycle of freezing and thawing was performed for a given set of samples and then repeated with new volumes of distilled water. We do not report error bars since the initial TC temperature at which the experiment is launched may slightly vary from one trial to another. Instead, we relied on the experiment's repeatability to extract trends and their margins of error.

**Data buildup**

The different variables we examined are presented in Figure 2a for samples frozen at RT (typically 20°C). When introduced into the freezer set at -30°C, the water in the cups starts cooling until the onset of freezing (OF, in minutes) occurs at 0°C, which is taken as the initial information. The time from initial temperature to onset gives the cooling rate (in °C/min), which we will also discuss. Then, freezing occurs until all the ice is formed, and the temperature decreases again. We record the time at which the sample temperature shifts by approximately -0.5°C, taken as the end of the freezing point (EF, in minutes). The total freezing time (TF, in minutes) is TF = EF - OF.

Figure 2b first shows the onset of freezing and the cooling rate of all tested samples against an RT cup (data point at 20°C, thus corresponding to blank experiments). The initial temperature was varied between 2.5°C and 80°C. The onset of freezing logically increases with the initial temperature but somehow levels above 40°C. This is because the cooling rate increases when the temperature difference between the cup and the freezer is high. Note that the cooling rate increases linearly with temperature throughout the data, except when the temperature of TC is low (below 10°C) (Figure 2b). This contradicts several papers that have argued that the hot cup would start freezing before the cold one (e.g.[21]), albeit when avoiding supercooling. We also report in Figure 2b the onset of freezing of the reference cup, which was set initially at about 20°C.

The onset of freezing is constant in most TC temperature ranges and equal to the one at which both cups are first set at approximately 20°C. We can notice, however, a significant decrease in the onset of freezing at both extremes when the second cup is really cold (typically 2.5°C) or hot (typically 80°C). These data were replicated and are not subject to experimental biases.

Figure 2c shows the time of freezing of cups initially at various temperatures as measured experimentally, for which an asymmetric bell-like curve is observed (blue markers). The freezing time linearly decreases with increasing temperature for experiments with hotter samples compared to reference water. Despite some discrepancies between the values (we subtract two data points, onset, and end-of-freezing, both of which show slight uncertainties), a clear trend of a linear decrease in onset is significantly observed. For cups containing water at temperatures below 20°C, the freezing time is longer than for blank experiments, but a similar decrease is observed at the lowest temperatures (i.e., 2.5°C). Figure 2c also shows the freezing time of the reference cups as a function of the initial temperature of the TC. One notices a slight variation in freezing time, with the fastest freezing occurring when the complementary cup is cold.

Figure 2d shows the difference in freezing time between the sample and the reference cups as a function of the TC initial temperature. The time difference among all samples is several minutes, ranging from the coldest to the hottest experiments. It is demonstrated again here that the hottest sample in the device freezes faster than the coldest one, regardless of the initial temperature differences considered (within this experimental frame). As mentioned earlier, the significant error seen for the hottest samples is likely due to some unavoidable supercooling effect.

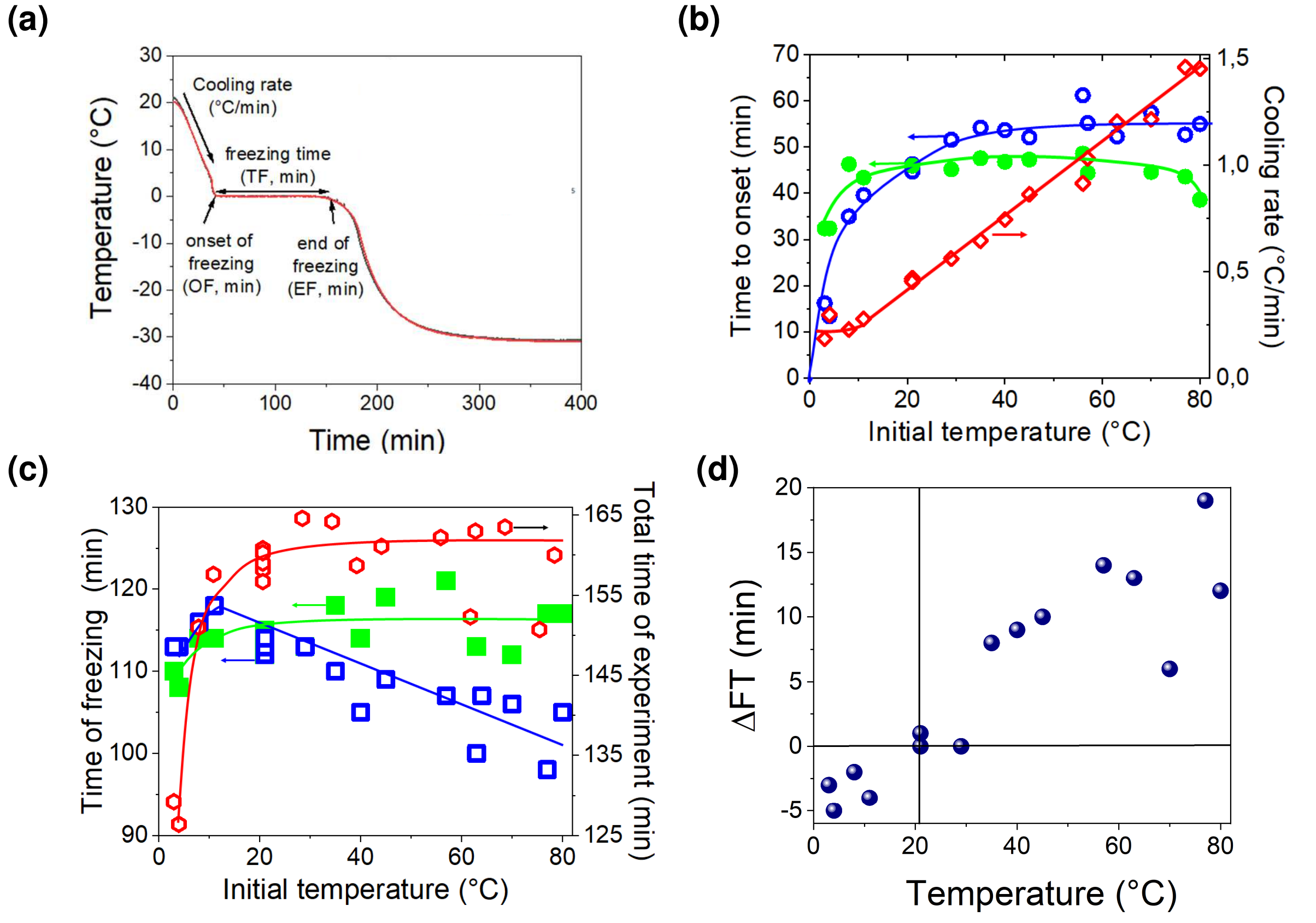

**Figure 2 – Analyses of the freezing curves.** (a) Typical parameters studied in this article. From the curves, the cooling rate, the onset of freezing (OF, time for which the thermocouple reaches 0°C), and the time of freezing (time between OF and thermocouple reaching -0.5°C) are extracted; (b) Influence of initial temperature of water on cooling rate (◇) and times to onset of the TC (○) and RC (●) cups; (c) Influence of the initial temperature of water on the total time of freezing (○) and times of freezing of the TC (□) and RC (■) cups; (d) difference in the times of freezing versus the initial temperature of the TC. Lines are only guides for the eyes.

**Heat transfer assessment**

At this point, the most probable explanation for the observed trends is to be found in heat transfers occurring in these experiments. First, to monitor the temperature evolution inside the freezer, we inserted a probe approximately 30 cm below the cups and another one in an empty cup (Figure 3a). The filled cup was first set at 81.4°C before being placed in the freezer and cooled. The first observation is that freezing the sole cup is faster than when the freezer is loaded with two filled cups, thus with twice the water content (42 g in each cup versus 42 g in one cup, as shown in Figure 2a). The probe in the freezer set at -30°C first slightly rises to -26.4°C and then quickly returns to -30°C. This variation, which occurs at the opening of the freezer (when introducing the hot cup), has been systematically observed in an equivalent blank experiment (two cups starting at a hot temperature, Figure 3b). Additionally, the probe inside the empty cup quickly decreases in temperature and then experiences a sudden increase, whereas the filled cup remains frozen. This is typical of a transfer of heat from the filled cup towards the empty cup. This transfer does not start automatically once freezing begins; instead, it stops once the process is complete. The temperature rise is relatively small (less than 1°C) but significant.

Heat transfer shall occur for any liquids other than water-based ones. To confirm this, we froze tetradecane, a non-polar solvent with a freezing point of 5°C. Here, experiments were rendered simpler since supercooling does not occur. Blank experiments on two RTs and two hot cups of tetradecane froze in concert and averaged to 81 and 83 min (4 trials each, std dev = 1.5 and 0.5, respectively). Then, we observe the same Mpemba effect when freezing hot (at 80°C) versus RT cups (see typical curves in Figure 3c). The freezing time across eight trials is 76.4 min (standard deviation = 3.0 min) for the hot cup versus 81.8 min (standard deviation = 2.0 min) for the RT cup.

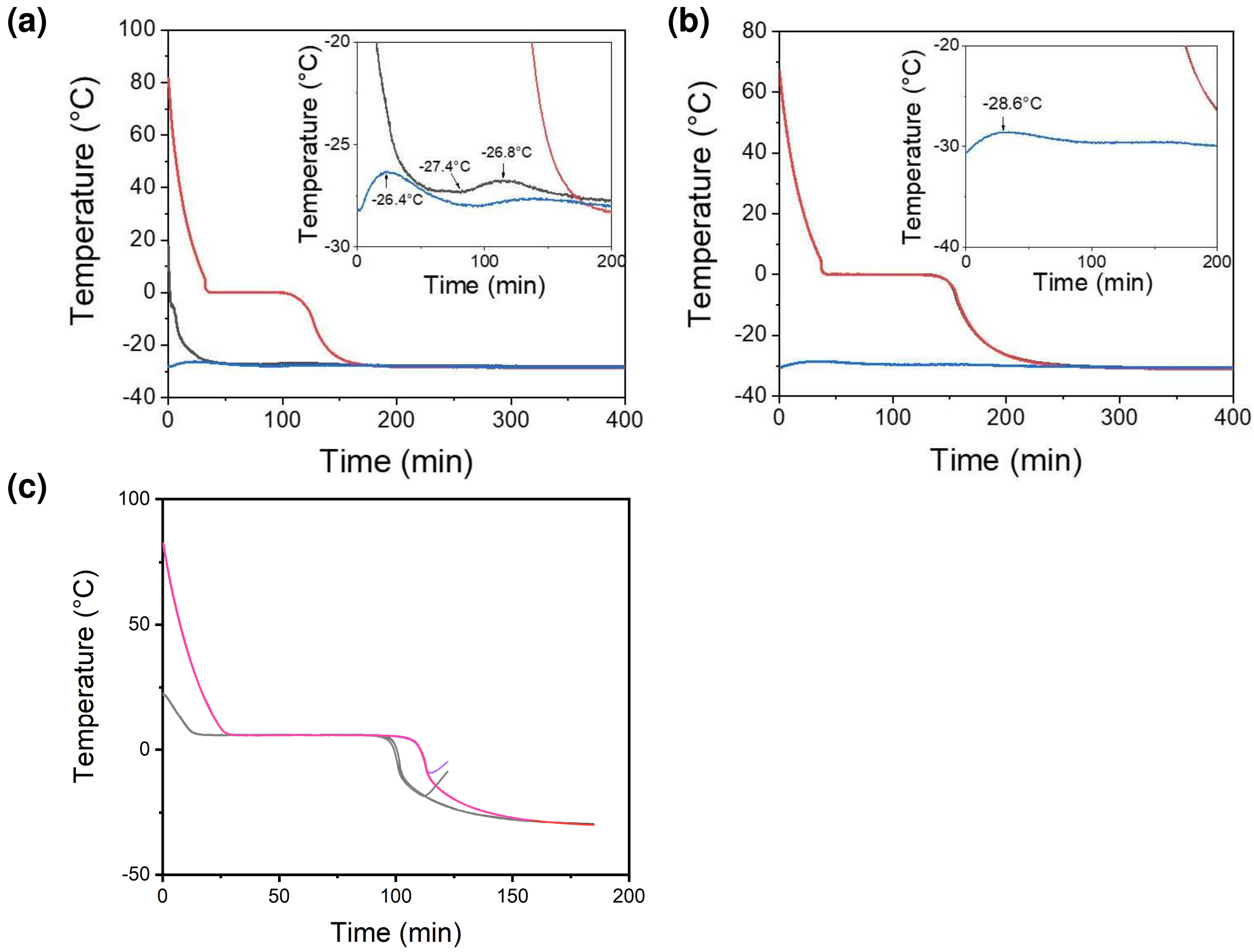

**Figure 3 – Complementary experiments**. Trials were carried out with one temperature probe at the back of the freezer in open air (in blue) with **(a)** one probe in a hot water-filled cup (in red), one probe in an empty cup (in grey), and **(b)** two probes in initially hot water-filled cups (in red and black). Insets: zooms on the low-temperature range. **(c)** Three tetradecane trials, including one that was prematurely stopped. The Mpemba effect is also seen here.

## SIMULATION

To complement the analysis of the trends derived from these experimental results, we present a simple numerical model that describes the freezing process of two identical volumes of water in two identical containers placed within an enclosure filled with air at atmospheric pressure, where the wall temperatures are controlled. One of the most interesting simulation outputs is the frozen fraction field at any given time for each cup. Integrating this field enables tracking the overall frozen fraction in each cup, thereby providing an exact estimation of the onset of freezing (when the frozen fraction exceeds 0) and the end of freezing (when the frozen fraction reaches 1). This method is more straightforward than estimating the temperature evolution at a specific point, as in the experimental study. Furthermore, simulation allows calculating energy-related quantities, such as the heat flux exchanged between the water and its environment during freezing. It provides additional insights and a better understanding of the causes behind the observed results. The model is implemented and solved using the finite element method from the commercial software COMSOL Multiphysics.

**Adapted geometry for simplified calculations**

The model geometry is illustrated in Figure 4a. All dimensions are detailed in Table S1 in the Supporting Information. While the dimensions of the cups and water volumes match those used in the experimental study, a fictitious 2D axisymmetric geometry was implemented to avoid the computational cost associated with a full 3D model. To do so, in the model the revolution axes of the two frustoconical cups are aligned. Furthermore, the arrangement of the cups within the chamber was chosen to be perfectly symmetrical, i.e., the free surfaces of the water volumes are positioned facing each other. If the test and reference cups have identical initial temperatures, their temperature and frozen fraction field evolutions will be rigorously similar. For different initial temperatures, any deviation in freezing time can be attributed solely to heat transfer phenomena between the two cups and their surroundings rather than to a geometric asymmetry effect. In the subsequent sections of this work, cup 1 will be referred to as the "test cup" and cup 2 as the "reference cup." Given the previously mentioned symmetry assumptions, this choice is entirely arbitrary and does not influence the results.

**Frozen fraction calculation**

The local frozen fraction $\varphi$ is defined as a function of the local temperature using a smoothed Heaviside step function:

$$\begin{cases} \varphi(T) = 0 \quad if\ T_f + \frac{\Delta T_f}{2} \leq T \\ \varphi(T) = 0.5 - 1.5\left(\frac{T - T_f}{\Delta T_f}\right) + 2\left(\frac{T - T_f}{\Delta T_f}\right)^3 \quad if\ \ T_f - \frac{\Delta T_f}{2} < T < T_f + \frac{\Delta T_f}{2} \\ \varphi(T) = 1 \quad if\ T \leq T_f - \frac{\Delta T_f}{2} \end{cases} \tag{1}$$

where $T_f$ is the freezing temperature of water and $\Delta T_f$ is the transition smoothing temperature range necessary for avoiding discontinuity and non-convergence in the numerical simulation. The model parameters related to the liquid-solid transition are summarized in Table S2.

From the local frozen fraction field, the overall frozen fraction $\Phi_i$ of the water volume in cup i (test or ref.) at time t can be calculated by:

$$\Phi_i(t) = \frac{1}{V_{w,i}} \iiint_{\Omega_i} \varphi\big(T(M,t)\big)\, dV \tag{2}$$

where $T(M,t)$ is the local temperature at time t, $\Omega_i$ denotes the water domain in cup i and $V_{w,i}$ is the volume of $\Omega_i$. Subsequently, the onset time of freezing of cup i, $t_{onset,i}$, is defined as the time for which $\Phi_i$ reaches the value 1x10-5, and the end time of freezing of cup i, $t_{end,i}$, is defined as the time for which $\Phi_i$ reaches the value (1 – 1 x 10-5).

### Heat transfer model and boundaries

In order to obtain the evolution of the temperature field $T(M,t)$ in the whole geometry, the time-dependent energy equation accounting for heat conduction is solved in the air domain, water domains, and cup wall domains:

$$\rho C_p \frac{\partial T}{\partial t} + \nabla \cdot (-k\nabla T) = Q_v \tag{3}$$

where $\rho$ (kg.m$^{-3}$) is the density of the medium, $C_p$ (J.kg$^{-1}$.K$^{-1}$) is the heat capacity, k (W.m$^{-1}$.K$^{-1}$) is the thermal conductivity. For water, ice, and air, these properties were treated as temperature-dependent, as detailed in the Supporting Information (Tables S2, S3, and S5) . Only the properties of the polypropylene cup walls were assumed constant (Table S4). $Q_v$ (W.m$^{-3}$) is a volume heat source reflecting the release of latent heat during the freezing process. Hence, in the water domains:

$$Q_v = \rho L_f \frac{\partial \varphi}{\partial t} \tag{4}$$

where $L_f$ (J.kg$^{-1}$) is the latent heat of melting/freezing and $\varphi$ is the average local frozen fraction (liquid: $0 < \varphi < 1$: frozen). In the air and cup wall domains, $Q_v = 0$.

Natural convection in air is not rigorously modeled, which would require coupling with fluid dynamics equations. Still, a simplified "convectively enhanced thermal conductivity" approach is considered (see supporting information). Radiative heat transfer is assumed to be negligible. At t = 0, it is assumed that the test cup, preheated uniformly at temperature $T_{test}$, and the reference cup, preheated uniformly at temperature $T_{ref}$, are instantly introduced into the cold enclosure at $T_{cool}$. The initial conditions are as follows: i) in the air domain: $T(t=0) = T_{cool}$; ii) in the test cup wall and water domains: $T(t = 0) = T_{test}$ and iii) in the reference cup wall and water domains: $T(t = 0) = T_{ref}$. On the external boundaries of the air domain, a constant temperature $T_{cool}$ is imposed during the entire simulation to represent perfect temperature regulation of the freezer walls. Along the symmetry axis, the standard no-flux boundary condition is applied. Perfect thermal contact (continuity of temperature and heat flux) is assumed at the air-cup wall, water-cup wall, and water-air interfaces.

**Freezing test reproduction**

Freezing simulations have been carried out with a fixed reference cup initial temperature $T_{ref}$ = 20.5 °C, a fixed enclosure temperature $T_{cool}$ = -30.5 °C and several test cup temperatures $T_{ref}$ = 2.5, 5, 10, 15, 20.5, 25, 30, 40, 50, 60, 70, 80, 90 °C, over a time span from 0 to 400 min. The numerical solution provides the temperature $T(M, t)$ and local frozen fraction $\varphi(M, t)$ fields throughout the entire geometry at all time steps. Integration post-processing allows determining the overall frozen fraction $\Phi_i(t)$ in each cup i at all time steps.

For the $T_{ref}$ = 60°C case, Figure 4b shows the evolution of the water temperature in each cup at three different locations along the revolution (z) axis: at a distance of 12 mm below the water free surface, at the center of the cup, and at the bottom of the cup, and the evolution of the overall frozen fraction in each cup. Due to the fictitious geometric arrangement chosen for our simulations, cooling is faster at the bottom of the cups than in the upper areas, where more heat accumulates in the neighboring air between the cups. As expected, the reference cup starts to freeze first (①). The inflexion point of the temperature curve at the bottom of the cups indicates that freezing starts in the bottom region of the cups.

Figure 4c illustrates the temperature field in and around the cups at the significant freezing moments: onset, half-freezing, and end, while Figure 4d depicts the local frozen fraction field corresponding to these moments. We confirm here that the first location to be reached by the 0°C isotherm line is indeed the bottom edge of each cup. At this point, the temperature field shows that, when the test cup starts to freeze (②), the surrounding environment is significantly colder than it was for the reference cup during its onset (①). Freezing progresses from the periphery and bottom to the center and top of the water volumes (Figure 4d). A very sharp decrease in temperature marks the overall end of freezing as no more latent heat is released (Figure 4b). In our configuration, the area located approximately 12 mm under the water surface is the last one to achieve complete freezing, as confirmed by the overall frozen fraction reaching one simultaneously (⑤,⑥). Finally, thermal equilibrium with the enclosure temperature is reached after approximately 300 minutes.

**Simulated data buildup**

The time to onset of both cups has been determined for each test temperature and is plotted in Figure 5a. The average cooling rate, calculated as the ratio of initial temperature (in °C) to the time to onset of freezing, is also reported. The onset time of the reference cup remains nearly constant, with a mean value of 25.1 minutes. A slight increase can be observed for the trials with the highest test cup temperatures, indicating that the heat released by the test cup somewhat affects the cooling of the reference cup, slowing it down. In contrast, the test cup's onset increases with higher initial temperatures, reaching 61 minutes for a starting temperature of 90°C. Although less pronounced, the same leveling trend was observed in the experimental measurements at higher initial temperatures. The larger the initial temperature difference, the more delayed the onset of freezing

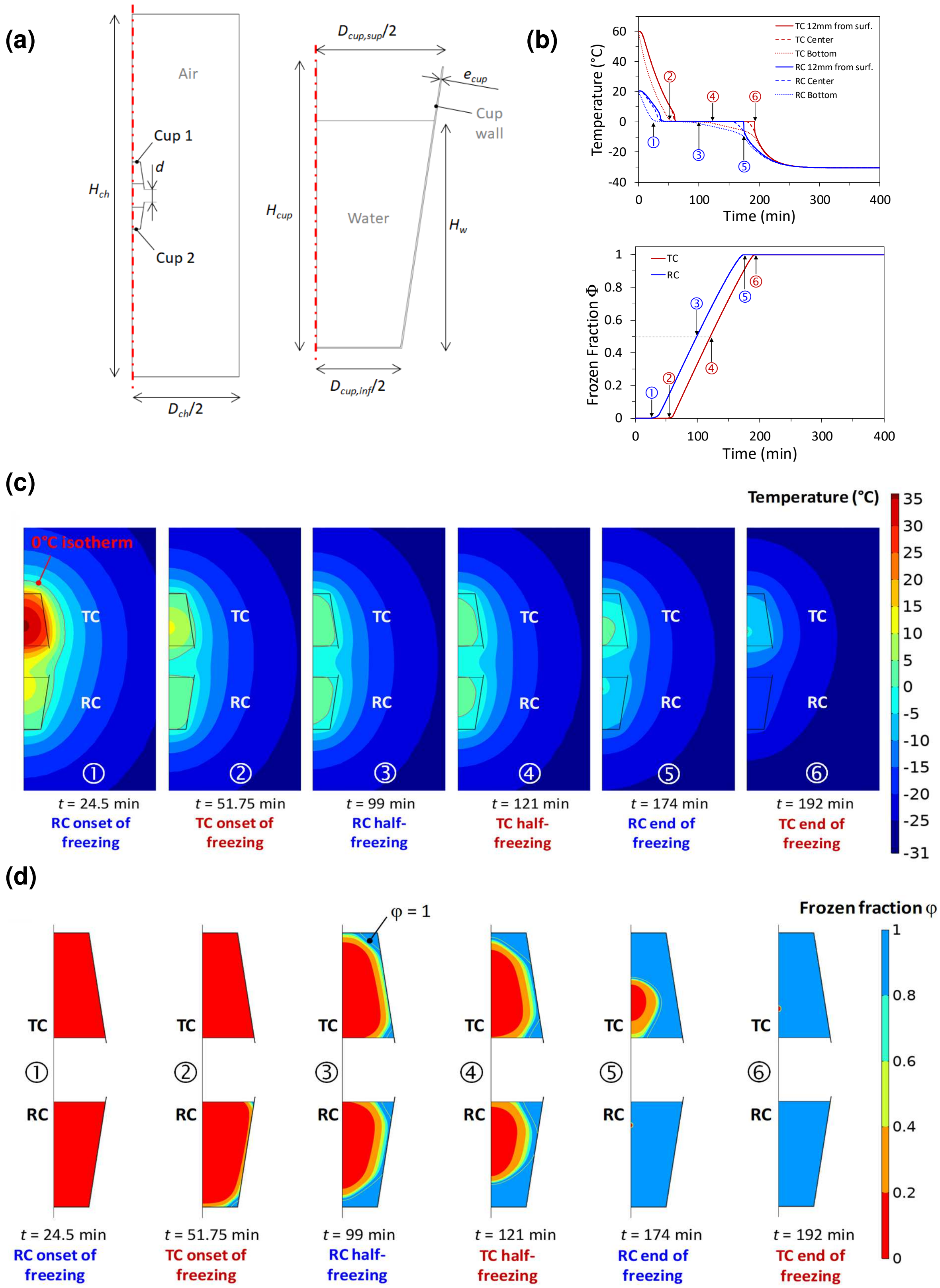

(a)
Air
Cup 1
d
Cup 2
H_ch
D_ch/2
D_cup,sup/2
e_cup
Cup wall
H_cup
Water
H_w
D_cup,inf/2
(b)
Temperature (°C)
Time (min)
TC 12mm from surf.
TC Center
TC Bottom
RC 12mm from surf.
RC Center
RC Bottom
Frozen Fraction Φ
TC
RC
(c)
Temperature (°C)
0°C isotherm
TC
RC
t = 24.5 min
RC onset of freezing
t = 51.75 min
TC onset of freezing
t = 99 min
RC half-freezing
t = 121 min
TC half-freezing
t = 174 min
RC end of freezing
t = 192 min
TC end of freezing
(d)
φ = 1
Frozen fraction φ
TC
RC
t = 24.5 min
RC onset of freezing
t = 51.75 min
TC onset of freezing
t = 99 min
RC half-freezing
t = 121 min
TC half-freezing
t = 174 min
RC end of freezing
t = 192 min
TC end of freezing

**Figure 4 – Simulation model and reproduction of the freezing tests (a)** Model geometry and dimensions with the overall view **(left)** and close-up view of a cup **(right)**; **(b)** Simulated cooling experiment for $T_{test}$ = 60 °C and $T_{ref}$ = 20.5 °C. **Top**: temperature evolution at three different locations along the revolution axis in test and reference cups; **bottom:** overall frozen fraction in test and reference cups as a function of time. Circled numbers indicate the significant times during freezing: ① RC onset - ② TC onset - ③ RC half-time - ④ TC half-time - ⑤ RC end - ⑥ TC end; **(c)** Evolution of the simulated temperature field during freezing of both cups for $T_{test}$ = 60°C and $T_{ref}$ = 20.5°C. The red curve is the 0°C isotherm; **(d)** Evolution of simulated local frozen fraction field during freezing of both cups for $T_{test}$ = 60°C and $T_{ref}$ = 20.5°C. The yellow line is the boundary of the fully frozen zone ($\varphi = 1$)

for the hotter cup compared to the colder cup. The average cooling rate logically increases quasi-linearly with the initial temperature, aligning with the basic principles of heat transfer physics, specifically Newton's law for convective heat transfer.

Note that from a theoretical perspective, the cooling rate of a body is proportional to the heat flux it loses to the surrounding fluid, which, in turn, is proportional to the temperature difference between the body's surface and the surrounding fluid. In the case of cooling by natural convection, the latter proportionality coefficient, namely the heat transfer coefficient, also increases as the body is hotter. This implies that the cooling rate should increase slightly faster than the linear trend when the temperature gap increases. This is evident in our simulation results, where the effect of natural convection has been incorporated into the model, albeit in a simplified manner (see the Supporting Information for details). However, despite a noticeable inflection, the simulation does not reproduce as sharply the experimental plateau in the onset of freezing time seen at high temperatures (Fig 2b). This is an expected consequence of the model's simplifications, as its approximated heat transfer model does not fully capture the other nonlinear heat transfer mechanisms (namely, radiation and, to some extent, evaporative cooling and mass loss), which all become more efficient at high temperatures.

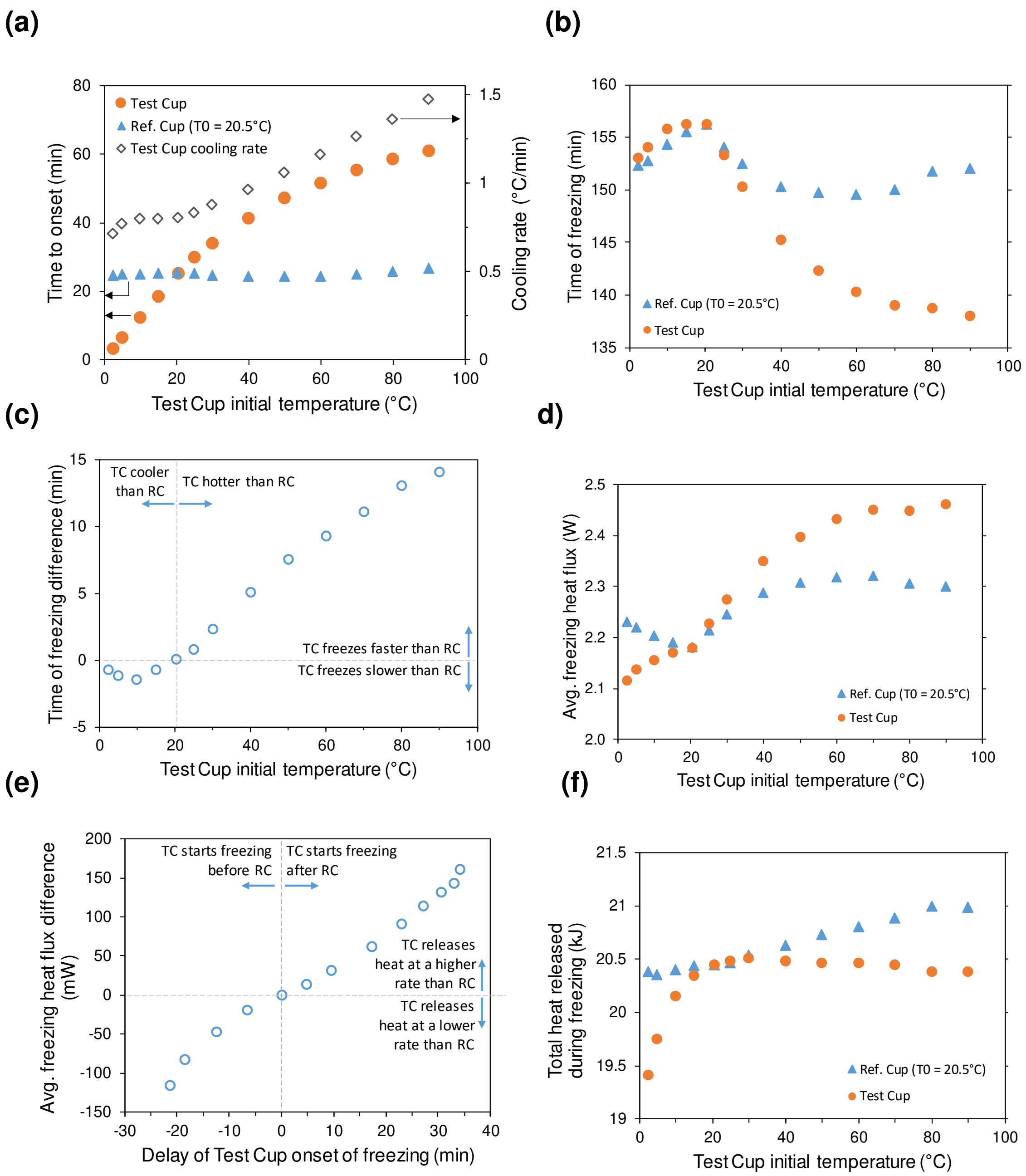

**Figure 5 – Results of the simulation.** Influence of initial test cup temperature on **(a)** times to onset of freezing and cooling rate, **(b)** times of freezing of test and reference cups, and **(c)** reduction in times of freezing of test cup versus reference cup; **(d)** Influence of initial test cup temperature on the average heat flux released during freezing; **(e)** Difference of the average heat fluxes released by the cups during freezing vs the freezing onset delay for the test cup; **(f)** Influence of initial test temperature on the total amount of heat released by both samples during freezing.

Figure 5b presents the freezing times of both cups, and Figure 5c shows the difference between these values and the initial temperature of the test cup. The values of the freezing time found are notably higher than in the experimental study. In this latter case, using temperature plots to detect the beginning and end of freezing results in an underestimation of the total freezing time, as it omits the initial and final stages of freezing. This discrepancy does not necessarily affect the value difference, as this bias is reproducible from one trial to another. The test cup's (and, to a lesser extent, the reference cup's) freezing time follows a bell-shaped curve similar to the experimental observations (Figure 5b vs Figure 2c). Indeed, the freezing time for both TC and RC is maximized when both cups have the same initial temperature, i.e., the blank trial. It should be noted that the blank trial at 20.5 °C exhibits precisely the same freezing time, indicating that the model is perfectly symmetric and the geometry introduces no bias.

The colder cup consistently requires more time to freeze than the hotter cup (Figure 5c). However, the difference decreases as the colder cup's initial temperature approaches 0 °C (typically below 10 °C). This reversing trend can be attributed to the rapid onset of freezing when the test cup is initially at a temperature close to freezing. In such cases, there is insufficient time for heat to transfer from the hotter reference cup to the surrounding air and ultimately reach the test cup before freezing begins. Consequently, the freezing of the test cup occurs in still-cold surrounding air, allowing for maximum heat flux, and as a result, its duration is not significantly affected. The order of magnitude of the time difference in freezing values closely aligns with the experimental results (see Figure 2d): a 60 °C initial temperature difference results in the hot test cup freezing approximately 13 minutes faster than the cold reference cup in the simulation.

**Digging deeper: heat exchanges and fluxes**

The average total heat flux (in W) released by the water contained in cup i (test or ref.) during the freezing transition (i.e., from $t_{onset,i}$ to $t_{end,i}$) can be calculated as:

$$Q'_{avg,i} = \frac{1}{t_{end,i} - t_{onset,i}} \int_{t_{onset,i}}^{t_{end,i}} \iint_{\partial\Omega_i} \overrightarrow{q"}(M,t) \cdot \vec{n} \, dS \, dt \qquad (5)$$

where $\partial\Omega_i$ denotes the boundary surface of the water domain in cup i (composed of the water-cup wall and water-air interfaces), $\overrightarrow{q"}(M,t)$ is the heat flux vector (W.m$^{-2}$) at any point M of $\partial\Omega_i$ at time t and $\vec{n}$ is the outward unit normal vector at point M. According to this definition, $Q'_{avg,i}$ is positive when the water releases heat to the surroundings, which is the case during freezing.

The calculated average heat flux values during freezing for reference and test cups are plotted against the test cup's initial temperature in Figure 5d. The average heat flux released during freezing by the test cup increases with its higher initial temperature. For the simulated case $T_{test}$ = 90 °C, the heat release rate of the test cup reaches 2.46 W, whereas the reference cup levels at 2.3 W. As shown previously (Figure 5a), the hotter cup initiates freezing later than the colder cup, given a larger initial temperature difference. In this scenario, when the hotter cup starts to freeze, the colder cup has already cooled to a relatively low temperature (around 0 °C) and transfers less heat to the surroundings, which are also cooler. According to Newton's law, this allows the hotter cup to release heat to its surroundings at a higher average rate, leading to a larger difference between the respective freezing heat fluxes of the hotter and colder cups, as shown in Figure 5e**.** Indeed, there is a nearly linear relationship between the time gap in the onsets of freezing and the difference in the average rate at which the cups release heat during freezing, which is typically directly correlated with the difference in freezing times. In the simulated cases, the maximum onset delay (34.25 min) is associated with a maximum heat flux difference of 160.7 mW, equivalent to a relative deviation of 7% compared to the reference flux.

Furthermore, the total amount of heat (in J) released by the water contained in a cup *i* during the freezing transition can be evaluated by:

$$Q_{tot,i} = Q'_{avg,i}\left(t_{end,i} - t_{onset,i}\right) \quad (6)$$

The corresponding values are plotted in Figure 5f. The total heat released by the test cup during freezing is nearly invariant, at approximately 20.5 kJ, for initial temperatures higher than that of the reference cup. Conversely, for initial temperatures approaching 0°C, this amount of energy decreases to 19.4 kJ. For the reference cup, the quantity of heat released during freezing remains relatively stable, varying between 20.4 and 21 kJ. From a physical standpoint, $Q_{tot,i}$ comprises two components: *i*) the latent heat required to freeze the water volume within the cup and *ii*) the sensible heat corresponding to the decrease in the average temperature of the water between the overall onset of freezing and the overall end of freezing. Since the cooling of the cups from their external surface is inherently non-uniform, a temperature gradient is building in the water, particularly in frozen regions where convection is absent. This non-uniform cooling causes freezing to initiate at the coolest point while some areas remain several degrees warmer. Similarly, freezing concludes at the warmest point while certain areas are already several degrees cooler. This can be visualized in the temperature field plots presented in Figure 4d. Therefore, the total heat released by the cups during complete freezing is not expected to be consistent across all experiments. While the latent heat component remains constant, the sensible heat component is influenced by the cooling kinetics, which are governed by the surrounding conditions. For a fixed mass $m_w$ = 53 g of water contained in one cup, the latent heat component corresponding to complete freezing is $Q_{lat} = m_w L_f$ = 17.7 kJ. Hence, the difference between the actual calculated value $Q_{tot}$ and the theoretical latent heat component $Q_{lat}$, ranging from 1.7 to 3.3 kJ represents sensible heat release that could be ascribed to the overall cooling of the cup during freezing.

## DISCUSSION

Here, we have proposed studying the influence of the initial temperatures of water cups simultaneously introduced into a freezer on their freezing process. We have established conditions where supercooling occurs rarely. We put the experimental and simulation results into perspective below.

On the experimental side, the rate at which temperature decreases when the cups are introduced in the freezer is consistent with thermodynamics (that is, with Newton's law of cooling): the hotter the sample, the larger the cooling rate. On the other hand, the onset of freezing rapidly plateaus for temperatures above 40°C, despite the increase in the temperature difference between the sample and the freezer air. The reference cup shows a constant time to onset, regardless of the initial temperature of the TC cup. At this point, it is unlikely that heat transfer occurs. The freezing time is maximum when both cups are set at the same temperature, regardless of its value. When introducing an initial temperature difference ΔT between RC and TC, either set to be positive or negative, the freezing time in TC is systematically less than that in RC. The time of freezing in RC is constant except when TC is at lower temperatures, in which case it is slightly less. A clear trend towards accelerated freezing of the TC cup compared to the RC cup when the absolute difference in temperature between RC and TC increases is obvious. The fact that the ‘hot’ water sample can transfer heat to the surrounding atmosphere explains such behavior: less time is needed to perform the exothermic freezing reaction.

The simulation qualitatively and quantitatively confirms this behavior using a relatively simple heat transfer model, even though the geometric setup is not entirely identical to the experimental one. The fact that this same conclusion is reached from two different geometries (the 3D 'side-by-side' experiment and the 2D 'face-to-face' simulation) strongly supports the generality of the proposed mechanism, proving it is not an artifact of a specific spatial arrangement. In addition, from a physical standpoint, the model deliberately neglects mass transfer (evaporation). This simplification was intentional: as our experimental data already ruled out evaporation as the primary explanation, the simulation serves as a key proof-of-concept. It allows us to confirm that the staggered heat transfer mechanism, resulting from the delayed onset of freezing, is on its own sufficient to explain the freezing time reduction effect. Freezing time is closely linked to the rate at which water can release a certain amount of heat. For initial temperature values of the test cup higher than that of the reference cup, the amount of heat released by the test cup during freezing remains approximately constant. This implies that the higher the rate at which this heat can be released (i.e., the heat flux), the shorter the freezing time. The same explanation applies to the reference cup, regardless of the temperature of the test cup. In contrast, for initial test cup temperatures approaching 0°C, the amount of sensible (and total) heat released by the freezing test cup decreases due to a lower temperature gradient. The difference in the heat flux released by each cup during freezing, given the same amount of heat to be released, explains the difference in freezing time. The heat flux is strongly linked to the temperature difference between the cooling cup and the ambient air, which temporarily warms in the vicinity of the cups. This effect is more pronounced during their initial cooling phase before freezing, occurring at relatively high temperatures. The longer the test cup is preheated, the later it will start freezing compared to the reference cup. Consequently, the freezing of the test cup occurs in an environment whose temperature has had time to decrease due to the thermal regulation of the cold chamber. Furthermore, when the reference cup has completed freezing, its temperature will rapidly decrease toward that of the regulated enclosure. This, in turn, enables the test cup to finish freezing in an even colder environment.

According to Newton's law, this results in a higher average heat flux and, therefore, a shorter freezing time. Indeed, we have observed that the difference in average heat flux is nearly linearly correlated with the delay with which the hotter cup begins to freeze compared to the colder cup. The most unfavorable case occurs when both cups start cooling from the same temperature (or very close temperatures): freezing begins simultaneously, meaning that both cups release latent heat to the surrounding air simultaneously and for the same duration. Hence, during the freezing phase, the temperature gradient between the cup surfaces and the surrounding air remains low, resulting in a limited heat flux and an extended freezing time.

Note that analyses of the Mpemba effect on the basis of heat transfer, evaporative cooling, and natural convection phenomena have already been proposed[37,38]. However, the authors claimed they were unable to convincingly reproduce the observed effect, perhaps due to supercooling that was not strictly controlled in their experiments. They suggested the temperature-dependence of the hydrogen bond dynamics may play a key role in the effect[39]. The present work clearly rules out this hypothesis, as the same effect is observed in a non-polar liquid such as tetradecane.

## CONCLUSION AND PERSPECTIVES

In conclusion, under clearly defined conditions (such as no supercooling and simultaneous freezing), the Mpemba effect can be experimentally reproduced and measured with standard temperature sensors. The explanation is simple when looking at heat transfer from one cup to another. It applies to water as well as to a non-polar solvent like tetradecane, showing that it is not linked to the specific physical structure of water. We expect these results to help clarify the long-standing debate about this counterintuitive Mpemba effect.

Still, several questions remain open that would require additional experiments, some of which we attempted to address in this study on frozen water. To start with, only distilled water and tetradecane were shown here. We have conducted numerous experiments with various water solutions, including highly concentrated water (from a pond near Binghamton) and Sodastreamed water. The water containing a large content of solutes showed a retarded onset of freezing and freezing time but showed similar freezing shape (Figure S6). On the other hand, adding $CO_2$ to the TC cup slightly accelerates the freezing (Figure S7). The difference is due in part to the disappearance of the maximum density effect in the gas-filled sample. Further investigation would be required to assess the impact of these different additives on the freezing process. In addition, we have observed that, initially, hot water samples would almost unavoidably (though slightly) supercool, for a reason we don't understand at this stage. This more generally questions the effect of boiling and freezing water on supercooling, a matter that one of us has recently addressed[40]. Moreover, we did not analyze the ice structure in each cup after it had frozen.

ASSOCIATED CONTENT

**Supporting Information**. The following file is available free of charge: Experiments showing the robustness of the freezing process, complementary main parameters, and approximation for the simulation and extra observations not interpreted in this work (PDF).

AUTHOR INFORMATION

**Corresponding Author**

Dr François Ganachaud. E-mail: francois.ganachaud@insa-lyon.fr

**Author Contributions**

The manuscript was written through the contributions of all authors. All authors have given approval to the final version of the manuscript.

ACKNOWLEDGMENT

F. Ganachaud thanks the CNRS for funding his sabbatical stay in COMPASS laboratory (2018-2019).

REFERENCES

[1]. M. Chaplin, "Water Structure and Science". Available at: https://water.lsbu.ac.uk/water/ (accessed 14/04/2025)

[2]. N. Agmon, H. J. Bakker, R. Kramer Campen, R. H. Henchman, P. Pohl, S. Roke, M. Thämer, A. Hassanali, “Protons and Hydroxide Ions in Aqueous Systems”. Chem. Rev. 2016, 116, 7642−7672.

[3]. O. Bjorneholm, M. H. Hansen, A. Hodgson, L.-M. Liu, D. T. Limmer, A. Michaelides, P. Pedevilla, J. Rossmeisl, H. Shen, G. Tocci et al., “Water at interfaces”. Chem. Rev. 2016, 116, 7698−7726.

[4]. H. I. Okur, “The Jones–Ray Effect Is Not Caused by Surface-Active Impurities”, J. Phys. Chem. Letters 2018, 9, 6739-6743.

[5]. T. T. Duignan, M. Peng, A. V. Nguyen, X. S. Zhao, M. D. Baer, C. J. Mundy, “Detecting the undetectable: The role of trace surfactant in the Jones-Ray effect”. J. Chem. Phys. 2018, 149, 194702.

[6]. e.g. K. P. Gregory, G. R. Elliott, H. Robertson, A. Kumar, E. J. Wanless, G. B. Webber, V. S. J. Craig, G. G. Andersson, A. J. Page, “Understanding specific ion effects and the Hofmeister series”. Phys. Chem. Chem. Phys. 2022, 24, 12682.

[7]. D. Lohse, X. Zhang, “Surface nanobubbles and nanodroplets”, Rev. Mod. Phys. 2015, 87, 981

[8]. M. Jeng, “The Mpemba effect: When can hot water freeze faster than cold?” Am. J. Phys. 2006, 74, 514−522.

[9]. E. B. Mpemba, D. G. Osborne, “Cool”. Phys. Edu. 1969, 4, 172−175.

[10]. I. Firth, “Cooler”. Phys. Edu. 1971, 6, 32−41. See also: e) ; f)

[11]. E. Deeson, “Cooler—lower down”. Phys. Edu. 1971, 6, 42−44.

[12]. D. G. Osborne, “Mind on Ice”. Phys. Edu. 1979, 14, 414−417

[13]. M. Freeman, “Cooler Still”. Phys. Edu. 1979, 14, 417−421.

[14]. G. S. Kell, “The Freezing of Hot and Cold Water”. Am. J. Phys. 1969, 37, 564−565

[15]. J. Walker, “The Amateur Scientist: Hot water freezes faster than cold water. Why does it do so?” Sci. Am. 1977, 237, 246−257.

[16]. D. Auerbach, “Supercooling and the Mpemba Effect: When Hot Water Freezes Faster Than Cold”. Am. J. Phys. 1995, 63, 882−885.

[17]. B. Wojciechowski, I. Owczarek, G. Bednarz, “Freezing of aqueous solutions containing gases”. Cryst. Res. Technol. 1998, 23, 843−848;

[18]. J. H. Thomas, Senior Independent Study Theses. Paper 82. Available in https://openworks.wooster.edu/independentstudy/82. Deposited in 2008 (accessed 2025-10-13)

[19]. J. I. Katz, “When hot water freezes before cold”. Am. J. Phys. 2009, 77, 27−29;

[20]. J. D. Brownridge, “When does hot water freeze faster than cold water? A search for the Mpemba effect”. Am. J. Phys. 2011, 79, 78−84

[21]. Z. Thang, W. Huang, Y. Zhang, Y. Liu, L. Zhao, “Direct observation of the Mpemba effect with water: probe the mysterious heat transfer”, Infomat, 2022, e12352.

[22]. H. C. Burridge, O. Hallstadius, “Observing the Mpemba effect with minimal bias and the value of the Mpemba effect to scientific outreach and engagement”, Proc. R. Soc. A. 2020, 476, 20190829.

[23]. H. C. Burridge, P. F. Linden, “Questioning the Mpemba effect: Hot water does not cool more quickly than cold”. Sci. Rep. 2016, 6, 37665; b)

[24]. W. B. Zimmermann, “In search of a Mpemba effect protocol: Some hot water does cool and freeze faster than cold”. Chem. Eng. Sci., 2022, 247, 117043.

[25]. Y. Tao, W. Zou, J. Jia, W. Li, D. Cremer, “Different Ways of Hydrogen Bonding in Water—Why Does Warm Water Freeze Faster than Cold Water?” J. Chem. Theory Comput. 2017, 13, 55−76.

[26]. Y. Huang, X. Zhang, Z. Ma, Y. Zhou, W. Zheng, J. Zhou, C. Q. Sun, “Hydrogen-bond relaxation dynamics: Resolving mysteries of water ice”. Coord. Chem. Rev. 2015, 285, 109−165.

[27]. A. Kumar, J. Bechhoefer, “Exponentially faster cooling in a colloidal system”. Nature, 2020, 584, 64−68.

[28]. D. C. Elton, P. D. Spencer, ”Pathological Water Science – Four Examples and What They Have in Common” in *Water in Biomechanical and Related Systems* (Springer, 2021), chap. 8.

[29]. G. Vali, “Repeatability and randomness in heterogeneous freezing nucleation”. Atmos. Chem. Phys. 2008, 8, 5017−5031.

[30]. A. A. Klimov, A. V. Finkelstein, The Mpemba Effect in Pure Water Has a Stochastic Origin. Experimental and Theoretical Resolution of the Paradox, 2025, https://doi.org/10.48550/arXiv.2508.05607. Accessed 11/10/2025.

[31]. C. Jarzynski, Nonequilibrium Equality for Free Energy Differences, Phys. Rev. Lett. 1997, 78, 2690-2693.

[32]. Zhiyue Lu, O. Raz, Nonequilibrium thermodynamics of the Markovian Mpemba effect and its inverse, PNAS, 2017, 114, 5083-5088.

[33]. A research in Scifinder typing ‘Mpemba effect’ keywords gives 191 occurrences, with an exponential growth of papers since 2019 (search on the 14/04/2025).

[34]. G. Teza, J. Bechhoefer, A. Lasanta, O. Raz, M. Vucelja. “Speedups in nonequilibrium thermal relaxation: Mpemba and related effects”, arXiv preprint, 2025, DOI: 10.48550/arXiv.2502.01758v3. Accessed 2025-10-13.

[35]. I. Prigogine, *Non-equilibrium statistical mechanics*, Interscience Publishers: NY, 1962;(b)

[36]. S. R. De Groot, P. Mazur, *Non-equilibrium Thermodynamics*, Dover Publications Inc., 2003.

[37]. Vynnycky, M., Kimura, S., “Can natural convection alone explain the Mpemba effect?”, International Journal of Heat and Mass Transfer, 2015, 80, 243-255.

[38]. Vynnycky, M., and Mitchell, S. L., Evaporative cooling and the Mpemba effect. Heat and mass transfer, 2010, 46, 881-890.

[39]. Jin, J., and Goddard III, W. A., Mechanisms underlying the Mpemba effect in water from molecular dynamics simulations. The Journal of Physical Chemistry C, 2015, 119, 2622-2629.

[40]. F. Ganachaud, “An Alternative Hypothesis on Enhanced Deep Supercooling of Water: Nucleator Inhibition via Bicarbonate Adsorption”. J. Phys. Chem. Lett., 2025, 16, 261−264.

TOC Graph

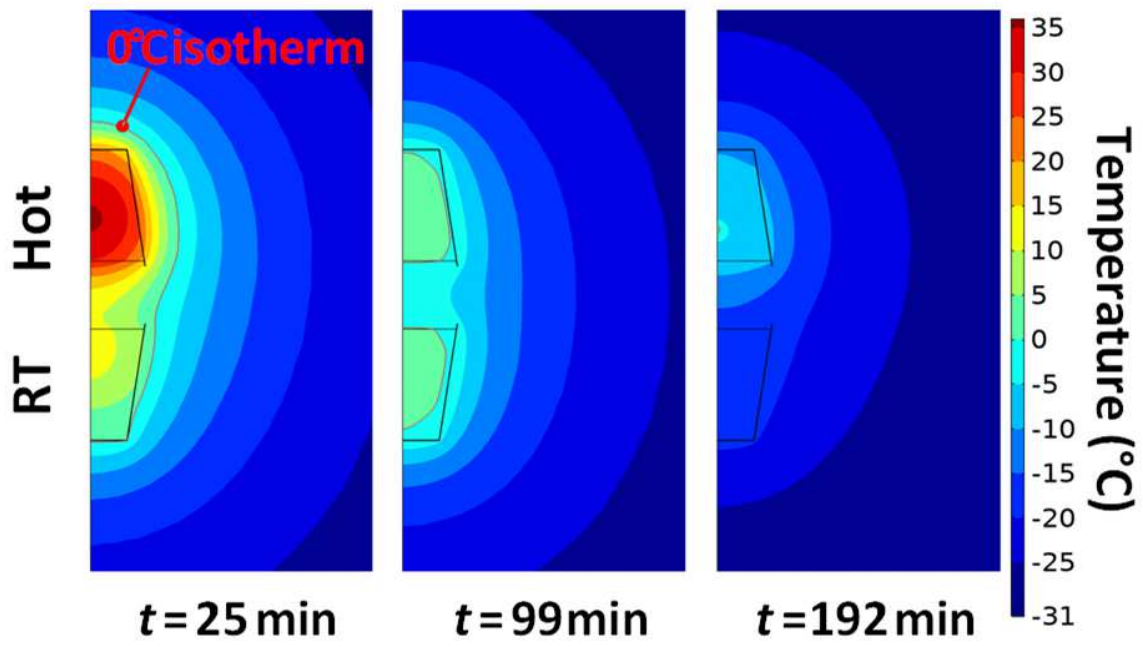

0°C isotherm
Hot
RT
t = 25 min
t = 99 min
t = 192 min
35
30
25
20
15
10
5
0
-5
-10
-15
-20
-25
-31
Temperature (°C)